\documentclass[10pt,a4paper,twocolumn,showpacs]{revtex4}
\usepackage{graphicx}
\usepackage{amsfonts}
\usepackage{amsbsy}
\newcommand{\beq}{\begin{equation}}
\newcommand{\eeq}[1]{\label{#1}\end{equation}}

\def\be{\begin{equation}}
\def\ee{\end{equation}}
\def\bea{\begin{eqnarray}}
\def\eea{\end{eqnarray}}

\begin{document}

\title{ Freeze Out and the Boltzmann Transport Equation}

\author{
L.P. Csernai$^{1,2}$, V.K. Magas$^3$, E. Moln\'ar$^1$, 
A. Nyiri$^1$ and K. Tamosiunas$^1$\\}

\affiliation{
$^1$ Section for Theoretical and Computational Physics,\\ 
     and Bergen Computational Physics Lab, BCCS-Unifob,\\
     University of Bergen, Allegaten 55, 5007 Bergen, Norway\\
$^2$ MTA-KFKI, Research Inst of Particle and Nuclear Physics,
     H-1525 Budapest 114, P.O.Box 49, Hungary\\
$^3$ Departamento de F\'{\i}sica Te\'orica and IFIC, Centro Mixto,\\ 
     Institutos de Investigaci\'on de Paterna - Universidad de Valencia-CSIC\\ 
     Apdo. correos 22085, 46071, Valencia, Spain\\
}

\begin{abstract}
{Abstract: Recently several works have appeared in the literature that addressed
the problem of Freeze Out in energetic heavy ion reaction and aimed for
a description based on the Boltzmann Transport Equation (BTE). In this
paper we develop a dynamical Freeze-Out description, starting from the
BTE, pointing out the basic limitations of the BTE approach, and the
points where the BTE approach should be modified.}
\end{abstract}


\maketitle

The Freeze Out (FO) is an important phase of dynamical reactions.
The connection of the kinetic description of this process and
the Boltzmann Transport Equation (BTE) raised considerable attention recently
\cite{E,B}

The phase-space (PS) distribution of post FO
particles can be obtained from kinetic FO calculations.
These were performed earlier  in
one-dimensional models \cite{FO1,FO3-HIP,FO4,FO5}, where the dynamics was
governed by two constants: re-thermalization and FO characteristic scales. 

The FO is a kinetic process and one would think it can be
handled perfectly by using the BTE, which
may describe both equilibrium and non-equilibrium processes 
in a 4-dimensional space-time (ST) volume element like
FO layer. This work and \cite{E} follows this approach. Our
aim is to analyse the situation, discuss the
applicability of BTE, and point out the physical causes, which
limit the applicability of the BTE for describing FO.

FO is usually assumed to happen on sharp 3-dimensional
ST hypersurfaces. However, the FO-fronts or FO-layers are not 
necessarily narrow, but they have
a characteristic direction (or normal, $d^3\sigma^\mu$), and it is more
realistic to assume a continuous, 4-volume FO in a layer (or domain)
of the ST. At the inside boundary of this layer
there are only interacting particles, while at the outside boundary hypersurface
all particles are frozen out and no interacting particles remain.
Then the ST volume element, $d^4x$, in the layer of interest
can be converted into
$d^4 x  \longrightarrow  ds \  d^3\sigma_\mu$,
where $ds$ is the length element in the direction of the 4-vector
$d^3\sigma^\mu$ ($\equiv d\sigma^\mu$ for short),  which can be
time-like or space-like.
Let us assume that the boundaries of this layer are
approximately parallel, and thus thickness of the
layer does not vary much.

The emission or FO probability may depend on physical
processes, cross-sections, transition rates, and the actual
PS distributions, $f(x,p)$.   Furthermore, $f$ must be 
determined {\it self-consistently} during the FO, i.e. all conservation laws should be
 satisfied across this transition 
layer and overall entropy should not decrease! \cite{Bu96,FO1,FO2,FO3-HIP,FO4,FO5,Cs04}.

\paragraph{Modified Boltzmann Transport Equation:}
We can derive the BTE from the conservation
of charges in a ST domain, $\Delta^4x$, assuming the
standard conditions \cite{LasBook}:
(i) only binary collisions are considered,
(ii) we assume "molecular chaos", i.e. that the number of
     binary collisions at position $x$ is proportional to
     $f(x,p_1) \times f(x,p_2)$, and that
(iii) $f(x,p)$ is a smoothly varying function compared to the
mean free path (m.f.p.). We have to
take into account that particles can scatter into the PS volume
element around $p$, or can scatter out from this volume element, 
described by Gain- and Loss- collision terms in the BTE.
In these terms
we consider elementary binary collisions where in the initial state
two particles collide with momenta
$p_1$ and $p_2$
into a final state of two particles with momenta
$p_3$ and $p_4$. In case of the Gain term the particle described by the
BTE, with momentum $p$ (without an index), is one from the two final
state particles, while in case of the loss term this particle is
one of the initial state particles. This is indicated by the
indexes of the invariant transition rate \cite{LasBook}.
We integrate over the momenta of the other three particles
participating in this binary collision. We use the notation
for the PS integrals:
$
{_{12}\mathcal{D}}_3 \equiv
\frac{d^3 p_1}{p_1^0}
\frac{d^3 p_2}{p_2^0}
\frac{d^3 p_3}{p_3^0} \ .
$
Shortening the notation further, by dropping the momentum arguments
of the PS distributions and keeping the indices only, yields:
\beq
p^\mu \partial_\mu f =
 \frac{1}{2}\int {_{12}\mathcal{D}}_4    f^i_1 f^i_2 W_{12}^{p4} 
-\frac{1}{2}\int {   _2\mathcal{D}}_{34} f^i   f^i_2 W_{p2}^{34}\ .
\eeq{BTE2}
Now, aiming for the description of FO let us
split up the distribution function, 
$f=f^i+f^f$, where $f^f$ is the PS distribution function
of the "free" or frozen out particles, 
while $f^i$ is the interacting component. \cite{FO1,E}
As the particles belonging to the free component may not collide any more,
they do not appear in the initial state components of collision integrals!

The gain term, $ f^i_1\ f^i_2\ W_{12}^{p4}$, populates both
the interacting, $f^i$, and free, $f^f$, components.
So, we introduce a FO probability, which 'feeds'
the free component,
${\mathcal{P}}_f  \equiv {\mathcal{P}}^{FO}(x,p)$,
and the rest, $(1-{\mathcal{P}}_f)$, feeds the interacting one.
Now, we can separate the two components into two equations.
The sum of these two equations returns the complete BTE above:
\bea \label{BTE1}
p^\mu \partial_\mu f^f &=&
 \frac{1}{2}\!\int\!\! {_{12}\mathcal{D}}_4 f^i_1 f^i_2 
{\mathcal{P}}_f W_{12}^{p4} \ , \\
p^\mu \partial_\mu f^i & =&
- \frac{1}{2}\int {_{12}\mathcal{D}}_4    
f^i_1 f^i_2 {\mathcal{P}}_f W_{12}^{p4}\\
&+& \frac{1}{2}\int {_{12}\mathcal{D}}_4    f^i_1 f^i_2 W_{12}^{p4} 
 - \frac{1}{2}\int {   _2\mathcal{D}}_{34} f^i   f^i_2 W_{p2}^{34} \ .
\nonumber
\eea
The free component does not have a loss term, because particles in the
free component cannot collide, and so, the free component cannot loose
particles due to collisions.
The first term of the second equation is a drain term, describing the escape or FO
of particles from the interacting component.
It is the inverse of the gain term for the free
component.
The last two terms are influencing the interacting component, and 
do not include the FO probability factors! Thus,
these two terms drive the interacting component towards re-thermalization.
For these the relaxation time approximation
can be used \cite{FO3-HIP,FO4,FO5}:
\beq
p^\mu \partial_\mu f^i=
-\frac{1}{2}\int {_{12}\mathcal{D}}_4 f^i_1 f^i_2 {\mathcal{P}}_f W_{12}^{p4}
\ +\ p^0\ \frac{f^i_{eq}-f^i}{\tau_{rel}} \ .
\eeq{BTEhalf2b}
We see that the structure of the kinetic equations,
used earlier \cite{FO1,FO3-HIP,FO4,FO5}, and the
separation of the "escape" and "re-thermalization" terms come
out in a simple straightforward way from the BTE.

However, the usual structure of the collision terms in the
BTE is not adequate for describing rapid FO, in a layer which is
comparable to the m.f.p. If we assume the existence of such a
layer this immediately
contradicts assumption (iii):
the change is not negligible in the direction of
$d\sigma^\nu$. The assumption of
"molecular chaos" is also violated in a FO process because
number of collisions is not proportional with
$f(x,p_1) \times f(x,p_2)$, but it is \underline{delocalized}
in the normal direction with
$f(x_1,p_1) \times f(x_2,p_2)$.
(The fact that the FO is a delocalized kinetic process, was
already used in ref. \cite{E} when integrals along the
path of propagating particles were introduced, but the consequences regarding
the details of the collision terms and the validity of the molecular chaos
assumption were not discussed.)

Based on the above considerations, one might conclude that the changes
of the distribution function are mediated by the transfer of particles, and
consequently only slowly propagating changes are possible. If the FO
layer propagates slowly, thus, its normal, $d\sigma^\mu$, must always 
be space-like.
This was a common misconception, where all "superluminous"
shock, detonation, deflagration fronts or discontinuities were considered
unphysical based on early studies \cite{Taub}.
However, it was shown recently, that discontinuous changes may
happen simultaneously in spatially neighbouring points, i.e. the
normal of the discontinuity-hypersurface can be time-like \cite{Cs87}.
This applies to the FO process also. Thus, the direction of characteristic
or dominant change,  $d\sigma^\mu$, may be both space-like and time-like
in the FO process.

From the all processes mentioned above (i.e. shocks, detonations, 
deflagrations etc.) 
the FO is the most special one, because the number of interacting 
particles is constantly 
decreasing as the FO proceeds, correspondingly the m.f.p. 
is increasing. In fact, it reaches infinity
when the complete FO is finished. 
This means that we 
can not make the FO in finite layer of
any thickness smooth enough to be modeled with the BTE. 
It is also obvious that if FO has some 
characteristic length scale,
it is not proportional with the m.f.p. only, because the m.f.p. increases 
as the density of interacting component becomes smaller, 
while the FO becomes faster in this limit, so its 
characteristic scale should decrease.

To describe that the PS distributions change rapidly
along the FO direction, we can introduce the 
Modified Boltzmann Transport Equation
(MBTE):
\bea \nonumber
p^\mu \partial_\mu f^f(x,p) &=&
\frac{1}{2}\int {_{12}\mathcal{D}}_4   {\mathcal{P}}_f W_{12}^{p4}\
f^i_1(x_1,p_1) f^i_2(x_2,p_2) \ , \\
\nonumber
p^\mu \partial_\mu f^i(x,p) &=&
-\frac{1}{2}\int {_{12}\mathcal{D}}_4  {\mathcal{P}}_f W_{12}^{p4}\
f^i_1(x_1,p_1) f^i_2(x_2,p_2) \\
& +& \  p^0\ \frac{f^i_{eq}-f^i}{\tau_{rel}} \ , \label{MBTE2}
\eea
where $x_k$ is the origin of colliding particles, i.e., the ST point
where the colliding particles were colliding last,
$x_k = x - u_k \tau_k$,  $\tau_k$ is the collision time,
$u_k^\mu = (\gamma_k, \gamma_k \vec{v}_k)$,
$\gamma = 1 / \sqrt{1-\vec{v}^2}$
and $\vec{v}_k = \vec{p}_k / p_k^0$. This is an essential modification
if the PS distribution has a large gradient in the ST.  This gradient
defines a ST 4-vector, characterizing the \underline{direction} of the process
$d\sigma^\mu$.
In ref. \cite{E} the direction $d\sigma^\mu$
is also introduced, however, it is not discussed why and it is
not connected to the delocalization of the BTE. 

The symmetries and the assumption of local
molecular chaos leads to the consequence that \underline{local} conservation
laws can be derived from the original BTE, i.e.
$\partial_\mu T^{\mu\nu} = 0$ and
$\partial_\mu N^\mu = 0$ were $T$ and $N$ are given as
momentum-integrals over the single particle PS distribution. 
After we delocalize the equations, this leads 
to new type of transport terms.
The FO process results in additional particle
and energy currents in the FO direction.

In the simplest case of infinitely
narrow FO layer (FO hypersurface) between $in$ and $out$ domains 
the distribution function is given 
by $f=f_{in}\Theta_{in}+f_{out}\Theta_{out}$ and common
BTE can be used everywhere except for the FO hypersurface. This case was
actually studied without introducing MBTE in ref. \cite{B} and the 
additional currents discussed above led to the
possibility of the new type of Phase Transition - the so called 
three flux discontinuity, see \cite{B} for more details.

The molecular chaos is also essential in the Boltzmann H-theorem. If we relax
this condition, the solution of MBTE does not converge to a stationary
thermal distribution, which enables us a simple thermo- and hydrodynamical
treatment of the physical problem.

\paragraph{Approximate Kinetic Freeze Out Models:}

Let us present the schematic derivation of the
kinetic FO model used by some of the authors earlier. This
represents only one particular possibility and the general
MBTE equation can be solved or approximated in other ways also.

If the ST distribution is non-uniform and the direction of steepest
gradient can be clearly identified, one may replace one (or
more) of the integrals over $d^3p_1$ (or $d^3p_2$) by ST
integrals over the origins of the incoming particle(s), $d^4x'$,
requiring that the particle reaches the ST point, $x$, when needed.
This requirement determines $p^\mu$ for a given $x'^\mu$.
It is reasonable to assume that after converting some of the
integrals to ST integrals and performing them, we get
an effective FO term reflecting the properties of 
local PS distribution, transition rate, and ST
configuration.

\begin{center}
\begin{figure}[ht]
\vspace{-0.5cm}
    \includegraphics[height=7.0cm, width = 8.4cm]{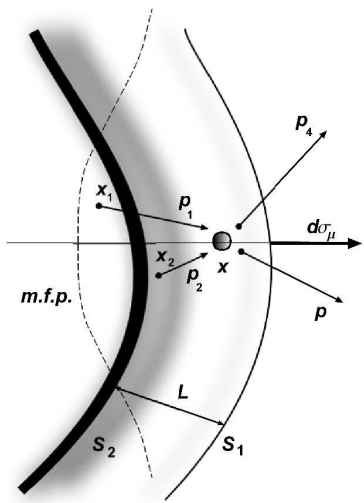}
\vspace{-0.5cm}
 \caption{\label{fig_mfp}
The plot of one of the  last collisions at $x$, 
shown in the spatial cross section 
of the FO layer. Particles arrive from positions $x_1$ and $x_2$ to point
$x$ with momenta $p_1$ and $p_2$. Within the FO layer of thickness, $L$,
the density of interacting particles gradually decreases (indicated by
shading) and disappears at the outside boundary, $S_1$ (thin line) of the
layer. R.h.s. from this boundary there are no interacting particles. 
Particles can reach $x$ from a region closer than the mean free path
(m.f.p. indicated by the dashed line), but only from places where the
interacting particle density is still not zero, i.e., mostly from the left.
The inside boundary of the FO layer,  $S_2$ (thick line) indicates the
points where the FO starts. Left of this line there is only interacting
matter and the FO probability is assumed to be zero for collisions happening
in the interacting region. 
}
\vspace{-1.1cm}
\end{figure}
\end{center}

Let us study the first equation of (\ref{MBTE2}), without
performing the integrals in a formal way, rather illustrating
the procedure giving a better insight to the problem.
When we are in the FO layer close to the boundary of complete
FO,  we have to calculate here the 
\underline{collision rate}.
According to the MBTE this depends on the PS distribution of the 
incoming particles at their origins, 
$ f^i_1(x_1,p_1) \times  f^i_2(x_2,p_2)$. 
Assume that the FO points in the direction $x$ 
(to the right), then on the r.h.s.
of the collision point density of particles is low or zero, while on
the l.h.s. it is larger, close to the pre FO value 
(see Figure \ref{fig_mfp}).
Most probably both incoming particles arrive from the left, where still 
there are many interacting particles and very few or none from the
right. Consequently most outgoing particles leave to the right.
Thus, the \underline{collision rate} at $x$ depends on the conditions
around $x - \lambda$, i.e. deeper inside the interactive
matter. So, the collision rate is  higher than the conditions
at $x$ could secure!
It does not go to zero even if we are at the
outside boundary of the FO layer, $S_1$, because particles still can arrive
from the left.

Let us execute two of the PS integrals for one incoming and one
outgoing particle,
$
\int \frac{d^3p_2}{p_2^0} \ \frac{d^3p_4}{p_4^0}\
$
\bea \nonumber
&\phantom{-}& p^\mu \partial_\mu f^f=
\frac{1}{2}\int {_{12}\mathcal{D}}_4  f^i_1 f^i_2 
                     {\mathcal{P}}_f W_{12}^{p4}  = \\
\label{BTEhf-4}
&=& \frac{Q_2 V_4}{2}  \int \frac{d^3p_1}{p_1^0}  f^i_1(x_1,p_1)  
               {\mathcal{P}}_f W_{1\bar{2}}^{p\bar{4}} \ ,
\eea
where 
$
Q_2     = \int \frac{d^3p_2}{p_2^0}\  f^i_2(x_2,p_2)
$ 
and  $V_4$
are invariant scalars. Eq. (\ref{BTEhf-4}) resembles eq. (3.27) in ref. \cite{LasBook}, but one of
the incoming particle distributions,  $f^i_2(x_2,p_2) $ is integrated out,
and leads to an integral quantity, $Q_2$. This can be approximated by the 
invariant scalar density at $\bar{x}_2$, i.e., 
$
Q_2 \approx n_2(\bar{x}_2) \ .
$
Here $V_4$ is not given explicitly, but can in principle be calculated
based on the distributions, $f^i_1(x_1,p_1)$ and  $f^i_2(x_2,p_2)$,
and the transition rate, $W$. The resulting transition rate
will then be averaged over particles 2 and 4: $W_{1\bar{2}}^{p\bar{4}}$.
So, finally we obtain:
\beq
p^\mu \partial_\mu f^f(x,p)=
\frac{Q_2 V_4}{2}  \int \frac{d^3p_1}{p_1^0}  f^i_1(x_1,p_1)
               {\mathcal{P}}_f W_{1\bar{2}}^{p\bar{4}} \ ,
\eeq{BTEhf-5}
where
$x_1=(t_1,\vec{x}_1) = $
$(t-\gamma_1 \tau_{coll.},\ \vec{x} - \vec{v}_1 \gamma_1 \tau_{coll.}) = $
$ x - u_1\ \tau_{coll.} $, \
(with the notations
$\vec{v}_1 = \vec{p}_1/p_1^0$ and
$u_1 = (\gamma_1, \gamma_1 \vec{v}_1)$)
indicates that the particle arrives at $x$
starting earlier from another point
$x_1$. As the first rough approximation we can take that 
$\tau_{cool.}$ is a mean collision time. The more general way to take into account particles 
coming from the different distances with corresponding propability factors is presented in \cite{praga04}.
In a rapid dynamical process the distributions at $x$ and $x_1$
are not the same, as it was discussed above.
Now, eq. (\ref{BTEhf-5}) can be integrated either in the $x_1$ -space,
or the $p_1$ -space. 

In addition, the FO probability, ${\mathcal{P}}_f$, may include
integrated information about the FO process, e.g. the probability
not-to-collide with anything on the way out. This should
depend on the integral number of interacting particles on the way
out.

For simplicity let us assume small angle scatterings, and the propagation
of a single particle
$W_{1\bar{2}}^{p\bar{4}} \approx $
$w_{\bar{2}}^{\bar{4}}\  \delta(p-p_1)$,
then
\beq
p^\mu \partial_\mu f^f(x,p)=
\frac{Q_2 V_4 }{2}
f^i(x_1, p)\ 
{\mathcal{P}}_f\  w_{\bar{2}}^{\bar{4}} \ ,
\eeq{BTEhf-6}
where $f^i(x_1, p)$ can be expressed in terms of $f^i(x, p)$ and
the 4-velocity, $u^\mu$, of the particle with momentum $\vec{p}$.

Let us now consider the FO situation, where we have a directed process
in a layer. The dominant change happens in the direction of the
normal of the FO hypersurface,
$d\sigma^\mu $ ($d\sigma^\mu \ d\sigma_\mu = \pm 1$)
and it is negligible along the hypersurface of the front. This actually means that
$
  p^\nu \partial_\nu f(x,p) \approx
 (p^\mu d\sigma_\mu)\ d\sigma^\nu \partial_\nu\ f(x,p)
$.
Inserting the above equation into (\ref{BTEhf-6}) yields to a kinetic equation
describing the directional derivative of the distribution function
in the direction of the dominant change, $d\sigma^\mu$, as
\beq
d\sigma^\mu \partial_\mu f^f(x,p) = f^i(x_1,p) \  P^*_{esc} \ ,
\eeq{MBTE7}
where $f^i(x_1,p)$ reflects the system properties one m. f. p. earlier 
than the collision point and the escape probability 
$
P^*_{esc} = P^*_{esc}(x,$ $p,$ $f^i,$ $d\sigma,$ $w,$ ${\mathcal{P}}_f)
$.
The derivation above did neglect several details and features, however,
reflects the basic structure of ad hoc kinetic FO models
\cite{FO1,FO3-HIP,FO4,FO5}.

\paragraph{Escape Probability:}

Here we just present briefly a direct
estimate for the escape probability \cite{magastalk}, detailed introduction and
analysis of which will be given in
\cite{Mo04a}.

The escape probability includes the collision rate and FO probability. The probability
not-to-collide with anything on the way out, reasonably should
depend on the number particles, which are in the way of a particle
moving outwards in the direction $\vec{p}/p$, across an FO layer 
of estimated thickness $L$ (representing the fact that we have finite number 
of particles on the way out to collide with \cite{FO3-HIP}). If we are in this FO layer in
 a position $x^\mu$, there is still
$\frac{L - x^\mu d\sigma_\mu}{\cos \Theta}$ distance ahead of us, where $\Theta$ is an 
angle between the normal vector
and $\vec{p}/p$. We assume then that
the FO probability is inversely proportional to some power of this 
quantity \cite{magastalk}. Thus
\be
   P^*_{esc} = 
   \frac{1}{\lambda(\bar{x}_1)}
   \left( \frac{L}{L-x^\mu d\sigma_\mu} \right)^a \left(\cos \Theta\right)^a   
   \Theta(p^\mu d\sigma_\mu)\,,
\label{esc1a}
\ee
where the power $a$ is influencing the FO profile across the front, and 
the cut factor is eliminating negative contributions to FO. 
In papers \cite{FO1,FO3-HIP,FO4,FO5} authors used $a=1$, 
and modeled FO in an infinite layer ($L\rightarrow \infty$), and 
a constant characteristic length $\lambda$ were used instead of  $\lambda(\bar{x}_1)$: 
\be
P^*_{esc} =\Theta(p^\mu d\sigma_\mu) \cos \Theta/\lambda\,.
\label{oldFP}
\ee
Comparing eqs. (\ref{esc1a}) and (\ref{oldFP}) one can see that we 
replaced the constant characteristic length $\lambda$, 
which was clearly oversimplifying the situation, with 
$
\lambda'(x) =\lambda(\bar{x}_1)\frac{L-x}{L}\ , 
$ (for $d\sigma^{\mu}=(0,1,0,0)$) which contains two factors.
The first is the collision rate, which is proportional with
$
\frac{1}{\lambda(\bar{x}_1)} \approx  \langle  n(\bar{x}_1) \sigma  \rangle \ ,
$
and this does not tend to zero even if we reach the outside
boundary of the FO layer, as this parameter is characteristic
to the interior region at $\bar{x}_1$.
The other is the generalized FO probability, which depends on the direction of the 
outgoing particle and on the number of interacting particles left in the
way to collide with, i.e. $\propto \frac{L}{L-x}$. 
Now the new characteristic length 
$\lambda'(x)$ 
gradually decreases as FO proceeds and the number of interacting particles decreases, 
and goes to $0$, when the FO is finished, 
as it was discussed above.

The simple angular factor, $\cos \Theta$, maximizes the FO 
probability of those particles, which propagate
in the direction closest to the normal of the layer, $ d\sigma_\mu$.
The quantities,
$\cos \Theta=p^x/| \vec{p} |$ for FO in $x$-direction and 
$\cos \Theta=1$ for FO in $t$-direction \cite{FO5},  
are not Lorentz invariant. Therefore to make our description 
completely invariant we shall generalize it to 
$\frac {p^\mu d\sigma_\mu}{p^\mu u_\mu} \sim \cos \Theta$ \cite{Cs04}.

So, we write the invariant escape probability, within the FO layer
covering both the timelike and space-like parts of the layer \cite{magastalk},
as
\be
   P^*_{esc} = 
   \frac{1}{\lambda(\bar{x}_1)}
   \left( \frac{L}{L-x^\mu d\sigma_\mu} \right)^a
   \left(\frac {p^\mu d\sigma_\mu}{p^\mu u_\mu}\right)^a\ 
   \Theta(p^\mu d\sigma_\mu)\,.
\label{esc1}
\ee

In refs. \cite{FO3-HIP,FO4} the post FO distribution was evaluated for
space-like gradual FO in a kinetic model with the escape probability (\ref{oldFP}). 
The post FO distribution depends on the details of the escape probability and
on the level of re-equilibration of the interacting component.
Bugaev assumed earlier \cite{Bu96} that the post FO distribution is a
(sharply) "Cut-Juttner" distribution, but the above mentioned model shows
that this can only be obtained if re-equilibration is not taking place.
The kinetic model provided an asymmetric but smooth PS distribution
\cite{FO3-HIP,FO4}, while the escape probability (\ref{esc1}) yields a somewhat
different, but also smooth PS distribution \cite{magastalk}. 

\paragraph{Conclusions:}
We have shown that the basic assumptions of BTE are not satisfied in the 
FO process, and so, the equation should be modified if used for the 
description of FO. We have also shown that earlier
ad hoc kinetic models of the FO process can be obtained
from our approach in a fully covariant way, and FO
in space-like and time-like directions can be handled on the same
covariant footing.

\vfill
\end{document}